\documentstyle[a4p,11pt]{article}

\def\hash{\symbol{"23}}
\def\sign(#1){(\!-\!1)^{#1}}
\def\binom(#1,#2){ (\!\!
     \begin{array}{c} #1 \\ #2 \end{array}\!\! ) }

\def\minus{\!-\!}

\def\Hpl{\hbox{H}}

\begin{document}

\thispagestyle{empty}
\title{
   \vskip-3cm{\baselineskip14pt
   \centerline{\normalsize\hfill NIKHEF-00-032}
  }
 \vskip.7cm
 {\Large\bf{New features of FORM}
 \vspace{1.5cm}
 }
}

\author{J.A.M. Vermaseren${}$
  \\[2em]
  {\it NIKHEF}\\
        {\it PObox 41882, 1098 DB Amsterdam, The Netherlands}
}

\date{}
\maketitle

\begin{abstract}
\noindent Version 3 of FORM is introduced. It contains many new features 
that are inspired by current developments in the methodology of 
computations in quantum field theory. A number of these features is 
discussed in combination with examples. In addition the distribution 
contains a number of general purpose packages. These are described shortly. 
\end{abstract}

\setcounter{page}{1}
\newpage 


\section{Introduction}

It is already more than 10 years ago that the symbolic manipulation program 
FORM was introduced. At the time it was constructed more or less as a 
successor to SCHOONSCHIP~\cite{schoonschip} which, due to the limitations 
of its language of implementation, became harder and harder to use. In 
addition the continuous development of methods of computation in many 
fields of science required a rapid expansion of the possibilities of the 
symbolic programs. Admittedly programs like Reduce, Maple and Mathematica 
offer a host of possibilities combined with extensive libraries, but the 
speed of SCHOONSCHIP was always something that set it apart from the other 
programs. Hence the task was to create a program that would be at least as 
fast as SCHOONSCHIP and be fully portable. In particular it should be 
usable on the fastest computers that are suitable for this kind of work 
(eg. vector processors will rarely give great benefits to symbolic 
manipulation in general. As we will see this is completely different for 
parallel processing). This resulted in version 1 of FORM. Due to more 
modern algorithms and the benefits of a second generation approach it is 
actually somewhat faster than SCHOONSCHIP when the problems remain limited 
in size and much faster when the expressions become large. Compared to the 
more popular computer algebra systems the difference in speed and use of 
memory depends largely on the type of the problem. This is due to a 
completely different philosopy of design and use. Typical factors are 
between 10 and 100, but there have been rare cases for which the factor was 
considerably less and cases for which the factor was considerably larger. 
Such a factor translates into a timeshift with respect to when a given 
calculation can be done in practise with other programs. Over the past 
years the computer industry has given us on average every three years a 
factor 4 in increase in capabilities. Hence for FORM this translates into a 
shift of more than 5 years. As a consequence most of the calculations in 
quantum field theory for which the available computer resources are a 
consideration have been using FORM over the past years.

Of course the development of FORM has continued. Version 2.0 was introduced 
in 1991. It contained a number of improvements that were inspired by the 
use of the program in a number of actual research projects. Since then 
regular small improvements were made, dictated by more and more complicated 
projects. After a number of years it became clear that a number of 
fundamental internal changes would enhance the capabilities very much and 
hence work on version 3 was started. Most of the code of version 2 was 
replaced, introducing better structured ways of dealing with internal 
objects. Even though the language of implementation is still regular C, the 
methods used are in many cases those of object oriented programming. It was 
of course important to do this in such a way that the speed and compactness 
of the expressions would not suffer from this. This has been achieved. 
After this revamping the new facilities were added. Some of them are 
completely new in the field of symbolic manipulation. This does not set the 
claim that such things are impossible with other programs. It just means 
that the FORM approach to some things is far more direct and often easier 
to program, even if some people may not like the language.

In this paper the most important new features are described shortly. In 
most cases some short examples of code are given with the corresponding 
output. It should be understood that this paper is not a manual. Because a 
number of packages comes with the FORM distribution, the examples to 
illustrate the features come mainly from these packages. In addition there 
are examples from some incomplete future packages. To improve the 
understanding of the above, first a very short description of FORM is given.

There is also a short section on a parallel version of FORM. It looks like 
this will be a major new feature that is currently still under development 
but results have been obtained already with one of the packages. The actual 
target is to run nearly unmodified FORM programs (the mincer package that 
has been used already was not modified at all) on a number of processors in 
such a way that to the user it looks like a much faster sequential machine. 
This is very useful for --among others-- the development cycle of big 
programs.

The final section discusses the availability of FORM. 
In the appendix we give a short description of the packages with some 
examples of their use. 

\section{A short overview of FORM}

Here a very brief overview of FORM is given. For a complete introduction 
one can consult the manual which contains a complete tutorial and a 
reference section.

FORM is not an interactive system. It runs in batch. In UNIX terms one 
would say that FORM is a filter. For large programs this is the natural 
mode anyway. It also allows the user to prepare programs with a familiar 
editor. The editor used by the author is stedi which has a folding 
capability that is recognized by FORM. The sources of this editor are 
available in the FORM distribution as well. If everan interactive version 
will be constructed, it will be based on this editor.

The language of FORM uses strong typing. This means that all objects that 
will be used in a program will have to be declared. This allows FORM to use 
a number of properties of such objcets and this adds to the speed of the 
program. Algebraic objects in FORM are expressions, symbols, functions, 
vectors and indices. In addition there are \$-expressions, sets, numbers, 
preprocessor variables and argument field wildcards. The functions are 
subdivided in categories of commuting and noncommuting functions. 
Additionally there are the classes of tensors, tables and regular 
functions.

The basic {\it modus operandi} of FORM is that variables are declared. Then one 
or more expressions are defined by giving them initial contents. After that 
operations are applied to one or more of these expressions.

A FORM program is subdivided in modules. Each module is an individual 
entity of which the contents are compiled when it is the turn of the module 
to be executed. After the module has been executed, the compiled version is 
forgotten again, but depending on the type of the module, declarations and 
expressions will be kept. Hence at the level of complete modules FORM acts 
as an interpreter. The contents of expressions as they were at the start of 
the module will be forgotten, unless the user takes some special action. 
This is needed to prevent the disk from filling up too quickly. At the 
beginning of each module extra variables can be declared and new 
expressions can be defined. If necessary old expressions can be deleted. In 
principle the operations of the module are applied to all old and new 
expressions.

The basic object for manipulation is the term. Expressions consist of 
terms, separated by addition operators (or subtraction of course). This 
addition operator is commutative and associative. Function arguments can 
consist of subexpressions, i.e. smaller expressions which are limited in 
size. Such subexpressions also consist of terms. Terms consist of subterms. 
The subterms are separated by the multiplication operator. This 
multiplication operator is associative and usually commutative. Only in the 
case of noncommuting functions and still unsubstituted objects that could 
contain noncommutative functions is the multiplication order sensitive. 
The number zero is not considered to be a term. An expression that is zero 
contains no terms.

FORM works its way through an expression in a sequential manner. It starts 
with the first term, applies all statements of a module to it and puts the 
results away for sorting (normal ordering of the complete expression is 
called sorting here. It does involve adding coefficients of terms that are 
identical with the exception of their prefactor). Partial sorting may occur 
when buffers become too full. After all terms of an expression have been 
treated this way the sorting is completed. Once all expressions have been 
treated this way the new versions of the expressions replace the old 
versions.

The actions inside a module can be either predefined operations or 
substitutions that may involve rather complicated patterns involving 
wildcarding. Wildcarding is the use of generic variables that represent a 
whole category of objects, both in the left hand side (the pattern) and in 
the right hand side (the replacement expression). 

A number of special functions is available to make the use of the program 
easier and often much faster. These involve amoung others a sum function, a 
number of combinatoric functions and a number of numerical functions. In 
addition there are special tensors like the Kronecker delta and the Levi 
Civita tensor. And there is much more.

For the use in particle physics there are Dirac gamma matrices and trace 
operations. It should be noted however that this is nowadays a rather 
unimportant part of the program as there exist many facilities to emulate 
these objects and operations (and similar operations that can be much more 
complicated).

Results of a calculation can be presented in many different ways. One way 
is to print them to regular output. They can be written to file in internal 
notation for use in future FORM programs. Or they can be written to special 
files, etc.

There is a powerful preprocessor that can manipulate the input at the 
character level. It is fully adapted to the requirements of symbolic 
manipulation. There are now also some mechanisms for the feedback of 
algebraic results to the preprocessor. This facilitates the possibility to 
have FORM programs write parts of themselves.

A number of buffers in FORM have to be allocated at startup. The user can 
influence these allocations with a special setup file. Ordinarily this is 
considered an advanced feature.


\section{New features}

For the description of the new features it is assumed that the user has 
some experience with the older versions. The distribution contains a good 
tutorial. The interested reader without experience could look at the many 
examples in it. Here we concentrate on the new concepts. For a complete 
description and the exact syntax and for the many additional new features 
we refer to the new manual.

\subsection{Variable administration}

The complete internal adminstration of variables has been made dynamic. 
This means that the old restrictions on the number of variables have been 
lifted. The setup file is no longer needed when more than 100 variables of 
the same type are asked for. The limit lies now at 6000 for 32-bit systems 
and $10^8$ for 64-bit systems.

Similarly the compiler buffer, the input buffer and the various 
preprocessor buffers are arranged dynamically. 
The reason that in the older versions these things were set up in a static 
way was due to the size of the computer memories in the late 80's.

\subsection{Preprocessor variables}

Preprocessor variables can be used `recursively'. The old notation to 
enclose a preprocessor variable in quotes has been changed into a notation 
in which the variable is enclosed in a backquote-quote pair as if these are 
brackets. This way they can be nested as in
\begin{verbatim}
#do itabs = 1,`size'
#ifndef `STABLE`itabs'HFILE'
#call table`itabs'
#endif
#enddo
\end{verbatim}
In the summer package there are 8 files table`itabs'.prc all having 
something like:
\begin{verbatim}
#procedure table5
#ifndef `STABLE5HFILE'
#define  STABLE5HFILE "1"
    lots of tables
#endif
#endprocedure
\end{verbatim}
This way the procedures are not called twice. Of course the contents would 
not be used twice anyway, but some of them are rather lengthy and FORM 
would have to read the whole procedure to find the $\hash$endif. That is 
avoided this way.

To make the preprocessor variables even more powerful 
there are also postincrement and postdecrement operators for 
preprocessor variables as in the C language:
\begin{verbatim}
     #define k "1"
     Local F`k++' = ...;
     Local F`k++' = ...;
     Local F`k++' = ...;
     Local F`k++' = ...;
\end{verbatim}
would result in the definitions of \verb:F1: to \verb:F4:. In the end 
\verb:k: would have the value 5. This is particularly handy when the 
program is generating expressions and we do not know in advance how many.
 
\subsection{Preprocessor abbreviations}

The notation has been cleaned up a bit by introducing a new operator for 
the compilation stages. This operator is indicated by three dots as in
\begin{verbatim}
    Symbols x1,...,x100;
    Local Fac10 = 1*...*10;
\end{verbatim}
Actually there is an even more general notation involving $<>$:
\begin{verbatim}
    id  f(<p1?,m1?,i1?>,...,<p6?,m6?,i6?>) = 
\end{verbatim}
Here the $<>$ enclose a pattern. FORM will look for the differences in the 
begin pattern and the end pattern, calculate their numerical differences 
and if they all have the same absolute value we can make a `running 
pattern' and make the substitution. Hence
\begin{verbatim}
    id  f(<p1,m4>,...,<p4,m1>) = 
\end{verbatim}
would evaluate into
\begin{verbatim}
    id  f(p1,m4,p2,m3,p3,m2,p4,m1) = 
\end{verbatim}
This facility cannot be used recursively but already it can clean up the 
code considerably when compared to how one would have to do this with do 
loops in the old versions of FORM. It becomes really powerful in 
combination with preprocessor variables:
\begin{verbatim}
    #do j = 1,`MAX'-1
        id  e_(i1?,...,i`j'?) =
                sum_(k,1,`MAX',e_(i1,...,i`j',k)*f(`j'+1,k));
        id  f(`j',k?) = tab(`j',k);
        .sort:step `j';
    #enddo
    id  e_(1,...,`MAX') = 1;
\end{verbatim}
This would be the central step in computing the determinant of a matrix 
that has been defined in the two dimensional table named tab. We use an 
intermediate function f to make sure that first the Levi-Civita tensors 
with two indentical indices are killed before table substitution is 
attempted. This is a very efficient algorithm for sparse matrices, 
especially when the matrices have been ordered in such a way that the first 
columns have the largest number of zeroes.

\subsection{\$variables}

These variables are at the same time small algebraic expressions and 
character strings. They can be used as either. When they are used and they 
are enclosed between a backquote-quote pair, the compiler will make their 
character representation and substitute this like a preprocessor variable. 
This use as a preprocessor variable even includes the possibility of using 
the postincrement or postdecrement operators. If they are not enclosed 
between the backquote-quote pair the contents are seen as an algebraic 
expression to be used during execution time. This can lead to very flexible 
code. If they are to be given an initial value, this can be done with the 
character $\hash$ before their name to have the preprocessor provide the 
value (hence this is done once during compilation time), or without the 
$\hash$ to indicate that this value should be provided during execution, 
each time a term passes this position in the code. There are additional 
methods to provide the \$-variables with a value that is the match of a 
wildcard variable. Let us have a look at an example:

\begin{verbatim}
    #$max = -100;
    if ( count(x,1) > $max ) $max = count_(x,1);
    .sort
    #do i = 1,`$max'
        ...
    #enddo
\end{verbatim}
Here we give \verb:$max: the initial value -100. During execution we look 
for the maximum power of the variable \verb:x:. Once the module has 
finished this maximum should be indeed the value of \verb:$max:, provided 
it is at least -100. At this point we can use this value as a parameter in 
a do loop. Note that the .sort is essential, because if we were to omit it, 
the do loop would have the bounds 1,-100 because the do loop instruction 
would be translated before any algebraic manipulation would have taken place 
(first the complete module is translated, then execution takes place).

It is also possible to manipulate the contents of the \$-variables, 
provided these do make sense.

The contents of these \$-variables are kept in memory. This means that the 
user should not try to see them as a substitute for regular expressions and 
have some of them grow to a large size. It would work, but performance 
might suffer when the physical memory is exhausted in the same way that 
Mathematica and Maple become rather disagreeable when they need more memory 
than is available.

\subsection{Hide}

When there are several expressions active at the same time, sometimes there 
is the need to temporarily put some of them `out of the way'. FORM has 
several mechanisms for this. The Skip statement would just skip the 
indicated expression(s) by copying them directly from the input to the 
output of the module. This is not always efficient. In addition one has to 
specify this in each module seperately. A more basic solution is to copy 
the expression(s) to a special location where they remain suspended until 
called back. This is done with the Hide and Unhide statements. The hidden 
expressions can still be used in the right hand side of expressions in the 
same way as fully active expressions can be used. This is a faster 
mechanism than the one that is used with stored expressions.
\begin{verbatim}
    .sort
    Hide F1,F2;
    Local F4 = F1+f3;
    .sort
         more modules
    .sort
    Unhide F1,F2;
    id  x = y+1;
\end{verbatim}
Here F1 and F2 are not operated upon for a few modules. Then they are 
called back and they are immediately active again. This means that the 
substitution of x also holds for the terms of F1 and F2.

There is another mechanism when one or a few statements should be applied 
only to a limited number of expressions. This concerns a new option of the 
if statement:
\begin{verbatim}
    if ( expression(G) == 0 );
       statements
    endif;
\end{verbatim}
In this case the statements are applied to the terms of all active 
expressions except for the terms of the expression G.

\subsection{$\hash$write}

This is a preprocessor controled instruction that can write either to 
regular output, to the log file or to any named file. It has a format 
string as in the printf function of the C language. It specifies what 
should be printed. There are some examples in which complete procedure 
files are constructed this way for later use in the same program.

\subsection{Print ""}

The new feature of the print statement is a mode like printf in the C 
language that can print information during execution. It can handle the 
printing of messages and even individual terms or \$-variables. This gives 
it great value as a debugging tool. A good example is given with the 
SplitArg statement in \ref{splitarg}.

\subsection{Table manipulations}

This facility involves a statement, a special function and a special print 
statement. The table\_ function can be used to load the contents of a table 
into an expression. The table indices are represented by powers of a given 
set of symbols. The fillexpression statement does the opposite. It takes an 
expression that is bracketed in a number of symbols and loads the contents 
into a table. The powers of the symbols indicate the table indices and the 
contents of the brackets form the corresponding table elements. The 
printtable statement writes a table to file in the form of fill statements. 
This file can be used in a future program to fill the table. Together these 
facilities allow the dynamic extension of tables in a program. If a table 
element that is needed has not yet been defined, one could compute it on 
the spot and add it to the table. The results can be written in such a way 
that in future programs the same element does not have to be recalculated.

\subsection{Symmetry properties}

Functions can have properties of symmetry. These properties hold for the 
set of all arguments. The functions or tensors can be symmetric, 
antisymmetric, cyclesymmetric and reversecyclesymmetric. This last means 
that the function is invariant both under cyclic permutations and a 
complete reversal of all arguments. These properties are put in the 
declaration of the function or tensor. Each time some arguments of the 
function are changed, FORM will check whether permutations have to be made. 
There are some restrictions. FORM will not accept argument field wildcards 
(wildcards that match zero, one or more function arguments) in patterns in 
which such a wildcard occurs inside a symmetric or antisymmetric function. 
The pattern matching would become extremely slow when no match occurs (all 
possibilities have to be searched and that involves n! possibilities for n 
arguments). There are tricks to circumvent this in most cases.

\subsection{Topology}

In combination with the symmetry properties FORM has now a command to look 
for loop structures in argument contractions. What this means is that if we 
have a network in which v is a vertex function and its arguments are 
indices and connections are represented by two vertices having identical 
indices, FORM can find the smallest loop in such a structure and make 
replacements based on this. There are some examples in the Color package in 
which the vertex function is the antisymmetric tensor f with three indices. 
In the program
\begin{verbatim}
#-
Indices i1,...,i9;
CF f(antisymmetric),ff;
Local F = f(i1,i2,i3)*f(i2,i4,i5)*f(i3,i5,i6)*
          f(i4,i7,i8)*f(i6,i7,i9)*f(i1,i8,i9);
ReplaceLoop,f,arguments=3,loopsize=all,outfun=ff;
Print +f;
.end
\end{verbatim}
the replaceloop command
takes the smallest loop, removes the loop indices and puts the remainder 
in the output function ff which should be at least cyclesymmetric (at least 
the action we take here implies it). If there 
is more than one possibility FORM takes the first one it encounters. The 
result of this program is
\begin{verbatim}
   F = f(i1,i8,i9)*f(i4,i7,i8)*f(i6,i7,i9)*ff(i1,i6,i4);
\end{verbatim}
Only a single loop has been replaced. If one would like to replace all 
loops one should put a repeat/endrepeat combination around the replaceloop 
statement. This would result in
\begin{verbatim}
   F = - ff(i1,i4,i6)*ff(i1,i6,i4);
\end{verbatim}

If such a command would not exist, the color package would be much more 
restricted because it would need a statement for each loop (loops with one 
vertex, two vertices, three vertices etc.) Also the corresponding pattern 
matching would be much slower. In the crucial phases of the program this 
made a factor 10 difference in speed and it also made the code much 
shorter.

\subsection{SplitArg}

\label{splitarg}
This is a rather useful tool for investigating arguments of functions. In 
its basic form it splits a multiterm argument into a number of arguments, 
each with a single term. There are variations for faster operation when 
only a single term should be taken from the multiterm argument. Take for 
instance the following example in which we want to select all functions den 
that contain the symbol \verb:j1: in its argument.
\begin{verbatim}
    SplitArg,((j1)),den; * take all terms in j1 separately
    id  den(j1) = den1(0,j1); * a single term does not get split
    *                           we may need more of these statements
    id  den(x?,x1?) = den1(x/x1*j1,j1)/x1*j1; * normalize the j1 argument.
    *
    * all den with two arguments have now just j1 without factor.
    *
    repeat id den1(x?,j1)*j1 = 1-x*den1(x,j1);
    repeat id den1(x1?!{x2?},j1)*den1(x2?!{x1?},j1) =
               (den1(x1,j1)-den1(x2,j1))*den(x2-x1);
    id  den(x?number_) = 1/x;
\end{verbatim}
Code of this nature occurs frequently in the summer package. It splits the 
fractions, being careful not to divide by zero (in some cases one has to be 
far more careful and involve theta\_ and delta\_ functions). Let us see 
what this code does. We use the following program (this is also a 
demonstration of the new print statement capabilities):
\begin{verbatim}
#-
S   j1,x,x1,x2;
CF  den,den1;
Off Statistics;
L   F = den(j1)*den(2+j1)*den(3-2*j1);
  Print +f "<1> %t";
SplitArg,((j1)),den;
  Print +f "<2> %t";
id  den(j1) = den1(0,j1);
  Print +f "<3> %t";
id  den(x?,x1?) = den1(x/x1*j1,j1)/x1*j1;
  Print +f "<4> %t";
repeat id den1(x?,j1)*j1 = 1-x*den1(x,j1);
  Print +f "<5> %t";
repeat;
    id den1(x1?!{x2?},j1)*den1(x2?!{x1?},j1) =
         (den1(x1,j1)-den1(x2,j1))*den(x2-x1);
    Print +f "<6> %t";
endrepeat;
id  den(x?number_) = 1/x;
  Print +f "<7> %t";
Print +f;
.end
\end{verbatim}
which results in
\begin{verbatim}
    #-
<1>  + den(j1)*den(2 + j1)*den(3 - 2*j1)
<2>  + den(2,j1)*den(3, - 2*j1)*den(j1)
<3>  + den(2,j1)*den(3, - 2*j1)*den1(0,j1)
<4>  - 1/2*den1(0,j1)*den1(2,j1)*den1( - 3/2,j1)
<5>  - 1/2*den1(0,j1)*den1(2,j1)*den1( - 3/2,j1)
<6>  - 1/2*den(2)*den1(0,j1)*den1( - 3/2,j1)
<6>  - 1/2*den(2)*den( - 3/2)*den1(0,j1)
<6>  - 1/2*den(2)*den( - 3/2)*den1(0,j1)
<7>  + 1/6*den1(0,j1)
<6>  + 1/2*den(2)*den( - 3/2)*den1( - 3/2,j1)
<6>  + 1/2*den(2)*den( - 3/2)*den1( - 3/2,j1)
<7>  - 1/6*den1( - 3/2,j1)
<6>  + 1/2*den(2)*den1(2,j1)*den1( - 3/2,j1)
<6>  + 1/2*den(2)*den( - 7/2)*den1(2,j1)
<6>  + 1/2*den(2)*den( - 7/2)*den1(2,j1)
<7>  - 1/14*den1(2,j1)
<6>  - 1/2*den(2)*den( - 7/2)*den1( - 3/2,j1)
<6>  - 1/2*den(2)*den( - 7/2)*den1( - 3/2,j1)
<7>  + 1/14*den1( - 3/2,j1)

   F =
       - 2/21*den1( - 3/2,j1) + 1/6*den1(0,j1) - 1/14*den1(2,j1);
\end{verbatim}
Note that the termwise option of the print statement gives us a trace of 
what happens and in what order.
 
\subsection{The term environment}

Sometimes one needs to work things out for just a few terms, but an 
intermediate sort for the results of these manipulations is needed in order 
to have things cancel. Or one would like to do even fancier things like 
trying whether the contents of certain brackets can be factorized by a 
given expression. If so, one would like to make that factorization. The 
term environment defines the current term temporarily as a small 
expression. Because this is done during the processing of the terms, 
sorting cannot be done with the .sort instruction which works for the whole 
expression. This means that we need a separate sort statement for all these 
small expressions. It should however be noted that the complete term 
environment consisting of the term statement, the endterm statement, and 
the intermediate regular statements and sort statements are all part of the 
same module and hence are all compiled before execution starts. Basically 
we use the same mechanism of operation in the argument environment. There 
function arguments can be operated upon and when the endargument statement 
is encountered the results are sorted and put back inside the function. The 
sort statement is hence equivalent to an endargument statement followed by 
a new argument statement. This last construction has the limitation that 
the result has `to fit' and hence is limited to MaxTermSize or less. The 
Term environment does not suffer from this restriction because we do not 
have to keep the results inside a function argument and also in the 
intermediate stages even the disk can be used and the expression may be 
quite large.
\begin{verbatim}
#-
S   x,y,[x+1],[x+2],x1,x2;
CF  acc;
L   F = y*(x+1)^2*(x+2)+y^2*(x+1)*x+y^3*(x+2)*x+y^4*x^2;
B   y;
Print +f;
.sort
Collect,acc;
Term;
    $min1 = 1000;
    $min2 = 1000;
    id acc(x?) = x;
    id  x = x1-1;
    sort;
    if ( count(x1,1) < $min1 ) $min1 = count_(x1,1);
    sort;
    Multiply ([x+1]/x1)^$min1;
    id  x1 = x+1;
    id  x = x2-2;
    sort;
    if ( count(x2,1) < $min2 ) $min2 = count_(x2,1);
    sort;
    Multiply ([x+2]/x2)^$min2;
    id  x2 = x+2;
EndTerm;
Print +f;
.end
\end{verbatim}
which gives
\begin{verbatim}
    #-

   F =
       + y * ( 2 + 5*x + 4*x^2 + x^3 )
       + y^2 * ( x + x^2 )
       + y^3 * ( 2*x + x^2 )
       + y^4 * ( x^2 );

   F =
      x*y^2*[x+1] + x*y^3*[x+2] + x^2*y^4 + y*[x+1]^2*[x+2];
\end{verbatim}
Of course this works only if one has a suspicion which factors are involved 
and things would go much better if FORM would be able to deal with 
factorizations by itself. This is considered for the future. For now this 
provides at least a partial solution.

\subsection{Bracket indexing}

In the old versions FORM would look for the contents of brackets inside an 
expression in a sequential manner. For relatively small expressions in 
computers with not much memory this was still not optimal, but also not a 
complete disaster. With the growing of the computers and hence the 
expressions the inefficiency of this method became a major factor as it is 
an algorithm that has a linear dependence of the size of the expression. 
Hence now the brackets can be indexed if the user asks for it. The index 
will of course take space of its own, and this space can approach (in 
principle) the size of the expression if the brackets have very little 
contents. This could result in bad problems again. Therefore there is a 
maximum to the size of the index (can be changed in the setup file) and 
when this maximum is reached FORM will start skipping brackets in the 
index. If such a skipped bracket is asked for, FORM will do a linear 
search, starting from the nearest bracket in front of it. The result is 
still a speedup by a big factor. The brackets in the index are located by a 
tree-like binary search which gives a logarithmic dependence of the number 
of brackets in the index. The degrading factor when brackets are skipped is 
proportional to the size of the expression divided by the number of 
brackets in the index. Hence in that case the improvement is proportional 
to a factor $B/\ln(B)$ with B the number of brackets in the index. Usually 
however there should be no problem in fitting the entire index inside the 
memory. Because this use of brackets is not very common and the making of 
the index does cost resources, the default is to not make an index. Here we 
have a simple example
\begin{verbatim}
#-
Symbols  x1,...,x10;
Local    F = (x1+...+x10)^10;
Bracket+ x1;
.sort
Drop F;
#do i = 0,10
Local    F`i' = F[x1^`i'];
#enddo
.end
\end{verbatim}
When we run this program the second module takes 1.59 sec on a pentium 300. 
Without the + in Bracket+ this step takes 2.84 seconds and this is not a 
very big expression yet. The program that solves the equations for the 
values in $N\rightarrow\infty$ in the summer package had an improvement 
factor of more than 100.

\subsection{Error messages}

The error messages have been changed completely. The most noticeable change 
is the indication of where the error occurs when the input listing is off. 
This mode is standard for procedures. There are additional warnings and 
there is a way to let FORM print its interpretation of the input. Together 
with the already existing printing of the name lists this provides extra 
tools for the debugging of FORM programs. In this example we have made two 
errors. It is the program of the bracket index. The name of the file is 
ex1.frm. Because we had turned off the 
listing, the line number information is rather crucial:
\begin{verbatim}
#-
Symbols  x1,...,x10;
Local    F = (x1+...+x10)^^10;
Bracket+ x1;
.sort
Drop F;
#do i = 0,10
Local    F`i' = FF[x1^`i'];
#enddo
.end
\end{verbatim}
When this program is run we obtain:
\begin{verbatim}
    #-
ex1.frm Line 3 --> Illegal position for operator: ^10
ex1.frm Line 8 --> Undeclared variable FF
++++Errors in Loop
\end{verbatim}

Because nearly all error messages have been changed and actually the 
complete compiler has been replaced, it is possible that some error 
messages either do not make complete sense or are too vague. The author is 
open for suggestions.


\section{The parallel version}

Because of the way FORM processes expressions it should be possible to 
achieve a high degree of parallelization for the complete program. After an 
initial study at the special computer at FNAL in 1991 this assumption was 
verified. It took however till the late 90's before a more serious attempt 
was made. This was made possible in the framework of a large phenomenology 
project centered at the university of Karlsruhe (DFG Contract Ku 502/8-1). 
Two people worked part time on this (Albert R\'etey and Denny Fliegner) and 
after struggling for a long time with the hardware and the parallelization 
library software they did indeed manage to obtain a working version of FORM 
that can run the Mincer package on an eight processor machine with a speed 
that ranges from 3 to 5 times the speed of a single processor. In addition 
many tests were made with other architectures~\cite{Parallel1}.

It should be noted that thus far the parallelization has been rather crude. 
The terms of the input expression in a module are distributed over the 
various processors. These do their work, and they sort their results. After 
this the outputs are collected by the master processor for a final merge 
and the output is written to disk. It is possible to be far more refined 
about this and make a much better load balancing by interfering, if needed, 
in the main expansion tree. The main problem however is the speed of 
transfer of operation. Thus far use has been made of the MPI protocol (and 
in tests of much more of course). In the future it is expected that the use 
of threads will increase the data transfer rate considerably. It is of 
course this transfer rate that puts a limit to the eventual speed that can 
be obtained. A given transfer speed in combination with slow processors 
will lead to saturation for a much larger number of processors than in the 
case that fast processors are used. In a shared memory machine with threads 
however the transfer rate would be mostly a function of the processor speed 
and hence much is expected of this model. Unfortunately the implementation 
is not yet complete.

At the moment the only readily available version is the one for a Compaq 
Alpha server (with 8 Alpha 21264 processors). It is expected that more will 
follow. A time prognosis cannot be given yet.

Of course it is not always possible to run code in parallel. Especially the 
new \$-variables could cause problems because their eventual contents 
depend on the order in which terms passed. If each processor keeps its own 
version of a \$-variable the processors may not have identical values at 
the end of the module. Take the simple code
\begin{verbatim}
#$max = -100;
if ( count(x,1) > $max ) $max = count_(x,1);
\end{verbatim}
which tries to find the maximum power of x. The various processors may have 
a different value in the end. Hence FORM should not parallelize modules in 
which \$-variables obtain values during running time, unless the user can 
provide some extra information. This can be done in the moduleoption 
statement. In that statement one can tell FORM that the final value of the 
\$-variable is either a sum, a maximum, a minimum or completely 
unimportant. In such cases FORM knows what to do with the results and the 
module can be run in parallel after all. It should be realized that this is 
all very new. Hence it could well be that many new facilities will be 
developed here and that there may be more ways in which future versions of 
FORM can decide about parallelization.

Any statement that concerns the parallelization will be completely ignored 
by a sequential version of FORM. This means that when a program is being 
prepared one could or should include already some of the indications for 
the parallel version. This way the upgrading to a parallel machine would be 
rather smooth.


\section{Availability}

Whereas version 2 of FORM has been commercial, it was judged that the 
scientific community as well as the reputation of the author would benefit 
more from a free distribution. Hence, starting version 3, FORM will be 
freely available again. For the moment this will involve a number of binary 
executables of the program. Amoung these will be at least executables for 
LINUX on PC architectures and for alpha processors running UNIX. More 
binaries will become available in the near future. 
The manual will be available both as a postscript file and as a .pdf file. 
This will allow the average user to obtain the system and print out its 
manual as well as view it on the computerscreen.

Of course it is hoped that this development will not only improve the 
visibility and popularity of FORM, but will also lead to more possibilities 
to improve FORM itself. The potential for this depends of course on people 
in charge of jobs knowing about the importance of FORM in the science 
community. Hence it is expected that people who use FORM for scientific 
publications will refer to the current paper. This way the use of FORM can 
be `measured'. The author hopes that the free distribution in combination 
with a proven popularity will lead to more people being involved in the 
development of FORM itself. There is still a whole list of potential 
improvements waiting for implementation. Undoubtedly many users will have 
their own suggestions. Also more packages would be appreciated.

The FORM distribution contains:
\begin{itemize}
\item Executables for a variety of computers.
\item The conv2to3 program with its sources. This program converts old 
FORM code into new FORM code.
\item A postscript and a pdf version of the manual and a tutorial by 
Andr\'e Heck.\item A number of packages. Currently these are `summer', 
`harmpol', `meltran', `color' and `mincer'.
\item The sources of the stedi editor. This editor works on BSD 4.0 systems 
and systems that still have an old compatibility mode. It also works under 
LINUX. There is a manual with it. Originally it was an editor for the Atari 
ST computers, made to work as well under MS-DOS. Then it was ported to BSD 
UNIX where it did run inside terminal windows. Under LINUX and X-windows it 
has recovered most of its old ease of use.
\item The sources of the minos database program for organizing large 
calculations. There is a small manual with it. The program should work on 
nearly any system. The database files should be system independent.
\item The axodraw system for including simple drawings and figures inside 
\LaTeX files. This includes the manual.
\end{itemize}

The files can be obtained from the www pages in 
http://www.nikhef.nl/$\sim$form.

\noindent Disclaimer: The files are provided without guarantee whatsoever. 
If some commands do not work as mentioned in the manual or as the user 
thinks they should work, the user should submit a bug report. This may or 
may not be reacted to depending on circumstances.

Acknowledgements: The author is thankful for many comments during the 
development of FORM. Many people contributed in the form of bug reports and 
useful suggestions. Some people however stand out. Denny Fliegner and 
Albert R\'etey for developing the parallel version of FORM (with support of 
DFG Contract Ku 502/8-1 ({\it DFG-Forschergruppe `Quantenfeldtheorie, 
Computeralgebra und Monte Carlo Simulationen'}), Geert Jan van Oldenborgh 
for being the main guinea pig during the early stages of the project, 
Andr\'e Heck for writing a good quality tutorial, Walter Hoogland for 
allowing me to spend so much time on the project in its early stages when 
success had not been proven yet.

\begin{appendix}
\section{The packages}

A few packages are provided with the FORM distribution. These packages are 
useful not only for their field of applicability. They also serve as 
examples of FORM packages and programming techniques. In this paper we took 
most of the examples from them. Here we give a very short description of 
what these packages are good for. More information is in a number of papers 
and in the future there may be some manuals.

\subsection{Summer}

This packages deals with harmonic sums. A full description of these sums is 
given in ref.~\cite{HarmonicSums}. We give a short overview.
The harmonic series is defined by
\begin{eqnarray}
    S_m(n)    & = & \sum_{i=1}^n \frac{1}{i^m} \\
    S_{-m}(n) & = & \sum_{i=1}^n \frac{\sign(i)}{i^m}
\end{eqnarray}
in which $m > 0$. One can define higher harmonic series by
\begin{eqnarray}
    S_{m,j_1,\cdots,j_p}(n) & = & \sum_{i=1}^n \frac{1}{i^m}
            S_{j_1,\cdots,j_p}(i) \\
    S_{-m,j_1,\cdots,j_p}(n) & = & \sum_{i=1}^n \frac{\sign(i)}{i^m}
            S_{j_1,\cdots,j_p}(i)
\end{eqnarray}
with the same conditions on $m$. The $m$ and the $j_i$ are referred to as 
the indices of the harmonic series. In the program we will use
$S_{j_1,\cdots,j_p}(n) \rightarrow $ \verb:S(R(j1,...,jp),n):.

There is an alternative notation in which the indices are one of the 
values 0,1,-1. A zero indicates that actually one should be added to the 
absolute value of the nonzero index to the right of the zero as in:
\begin{eqnarray}
    S_{0,1,0,0,-1,1}(n) & = & S_{2,-3,1}(n)
\end{eqnarray}

The weight of a sum is sum of the absolute value of its indices. In the 
notation with the 0,1,-1 it is equal to the number of indices. The harmonic 
sums with the same argument form a (weight preserving) algebra: The product 
of two sums with weights $w_1$ and $w_2$ respectively is a sum of terms, 
each with a single sum of weight $w_1+w_2$. The routine `basis' in the 
package implements this property. Example:
\begin{verbatim}
#-
#include summer6.h
.global
Local F = S(R(2,3),N)*S(R(-1,2),N);
#call basis(S)
Print +f;
.end
\end{verbatim}
This program gives the output
\begin{verbatim}
   F = - S(R(-3,2,3),N) - S(R(-3,3,2),N) + S(R(-3,5),N)
       + 2*S(R(-1,2,2,3),N) + S(R(-1,2,3,2),N) - S(R(-1,2,5),N)
       - S(R(-1,4,3),N) - S(R(2,-4,2),N) + S(R(2,-1,2,3),N)
       + S(R(2,-1,3,2),N) - S(R(2,-1,5),N) + S(R(2,3,-1,2),N);
\end{verbatim}
This defines a standard basis. A large number of sums involving harmonic 
sums and denominators can be done to any level of complexity. We give here 
a relatively simple example:
\begin{verbatim}
#-
#include summer6.h
.global
Local F = sum(j,1,N)*S(R(2,3),0,j)*S(R(-1,2),N,-j)*den(0,j);
#call summit()
Print +f;
.end
\end{verbatim}
which gives the output
\begin{verbatim}
Time =       0.15 sec    Generated terms =        104
                F        Terms in output =         78
                         Bytes used      =       2938

   F = - S(R(-1,-3,2,3),N) + S(R(-1,-2,-2,-1,3),N) + S(R(-1,-2,-1,-2,3),N)
       - S(R(-1,-2,2,-4),N) + S(R(-1,-2,2,-3,-1),N) + S(R(-1,-2,2,-2,-2),N)
                ....
       + S(R(1,2,-1,-3,-2),N) + S(R(1,2,-1,2,-3),N) + 2*S(R(1,2,-1,3,-2),N)
       + 3*S(R(1,2,-1,4,-1),N) - 4*S(R(1,2,-1,5),N) + S(R(1,2,3,-1,2),N);
\end{verbatim}
When $N\rightarrow\infty$ there are relations in addition to the regular 
algebraic relations. They allow all the sums in infinity to be expressed in 
terms of a rather small number of trancendental numbers of the Euler Zagier 
type~\cite{broadhurst}. In order to obtain these expressions one has to 
solve for each weight a number of equations. This number grows 
exponentially with the weight. The system has been solved exactly up to 
weight 9 and the file with the solutions is part of the distribution. It is 
about 20 Mbytes. In the file `summer6.h' only the solutions up to weight 6 
have been included. The programs that solve these systems of equations are 
also part of the distribution. In the case of weight 9 there were more than 
40000 equations in 13122 objects. The divergent sums were part of this 
because they can all be expressed properly into combinations of the basic 
divergence $S_1(\infty)$. This database of sums is rather important for 
whole categories of integrals as we will see in the harmpol and meltran 
packages. It should be noted that at the moment of writing, no constructive 
algerithms are known to determine these constants one by one. The only 
`algorithm' that comes close is the numerical evualuation of each 
individual sum, making an ansatz about which constants form a minimal set 
and then fitting all sums with a special numerical 
program~\cite{broadhurst}. This has been done for $w=10$ and in a 
restricted way for some higher values of $w$~\cite{broadhurstprivate}.

\subsection{Harmpol}
This is a package for Harmonic Polylogarithms as defined in 
ref.~\cite{HarmonicPoly}. A short description is in order here.

The weight one harmonic polylogarithms are given by:
\begin{eqnarray} 
  \Hpl(0;x) &=& \ln{x} \ ,          \nonumber\\ 
  \Hpl(1;x) &=& \int_0^x \frac{dx'}{1-x'} = - \ln(1-x) \ , \nonumber\\ 
  \Hpl(-1;x) &=& \int_0^x \frac{dx'}{1+x'} = \ln(1+x) \ . 
\end{eqnarray} 
For the higher weights we first define
\begin{eqnarray}
   f(0;x) &=& \frac{1}{x} \ , \nonumber\\
   f(1;x) &=& \frac{1}{1-x} \ , \nonumber\\
   f(-1;x) &=& \frac{1}{1+x} \ .
\end{eqnarray}
Then
\begin{equation}
\Hpl(\vec{0}_w;x) = \frac{1}{w!} \ln^w{x} \ ,
\end{equation}
with $\vec{0}_w$ an array with w zeroes while if the sequence $\vec{m}_w$ 
of w elements 0,1,-1 is not equal to $\vec{0}_w$
\begin{equation}
\Hpl(\vec{m}_w;x) = \int_0^x dx' \ f(m_w;x') \ \Hpl(\vec{m}_{w-1};x') \ .
\end{equation}
The weight of the harmonic polylogarithms is the number of indices. The 
argument $x$ can in principle take any complex value, but depending on the 
indices there may be a cut for real values $\ge 1$ or $\le 0$. These cuts 
are due to powers of $\ln(1-x)$ or $\ln(x)$. Such powers can be extracted, 
resulting in combinations of these logarithms and $\Hpl$-functions without 
these cuts. This will be needed for the Mellin transforms. It should be 
noted that we can use also the notation of the harmonic sums for the 
indices (with values beyond the set 1,0,-1).

Also the harmonic polylogarithms form a weight preserving algebra. In 
addition there are extra relations for $x=1$. One has to be rather careful 
with the divergent elements at $x=1$. The powerseries expansion of $\Hpl$ 
contains harmonic sums. The algebraic relations for general $x$ are related 
to the $N\rightarrow\infty$ relations of the harmonic sums and the extra 
relations in $x=1$ are related to the general algebraic relations for the 
harmonic sums.

The transformations $x\rightarrow 1-x$, $x\rightarrow 1/x$ and 
$x\rightarrow \frac{1-x}{1+x}$ give some interesting relations that allow 
for a reasonably fast numerical evaluation of the function over most of the 
complex plane (excepting the regions around $\pm i$).

The values in $x=1$ can be expressed in terms of the harmonic sums in 
$N\rightarrow\infty$. Because a weight w $\Hpl$-function is an integral 
over a denominator and a weight $w-1$ $\Hpl$-function this gives us whole 
categories of integrals that were till now extremely hard to solve. The 
results will be in terms of a limited number of trancendental numbers, but 
the relations for $x\rightarrow 1-x$ at $x=\frac{1}{2}$ will allow them to 
be expressed in terms of relatively fast converging series. Hence these 
numbers can be evaluated to any precision when the need arises.

Because the harmpol package uses a number of the routines of the summer 
package it will automatically load this package. Hence once harmpol.h is 
included, so is summer6.h. Example:
\begin{verbatim}
#-
#include harmpol.h
Off statistics;
.global
Local F = H(R(1,0,1),x)*H(R(-1,1,-1),x);
#call hbasis(H,x)
repeat id H(R(?a,n?!{1,0,-1},?b),x?) = H(R(?a,0,n-sig_(n),?b),x);
.sort
On statistics;
Print +f;
.end
\end{verbatim}
gives:
\begin{verbatim}
    #-

Time =       0.78 sec    Generated terms =         14
                F        Terms in output =         14
                         Bytes used      =        468

   F =
      H(R(-1,1,-1,1,0,1),x) + H(R(-1,1,0,1,-1,1),x) + 2*H(R(-1,1,0,1,1,-1),x)
       + 2*H(R(-1,1,1,-1,0,1),x) + 2*H(R(-1,1,1,0,-1,1),x) + 2*H(R(-1,1,1,0,1,
      -1),x) + H(R(1,-1,0,1,-1,1),x) + 2*H(R(1,-1,0,1,1,-1),x) + H(R(1,-1,1,-1
      ,0,1),x) + H(R(1,-1,1,0,-1,1),x) + H(R(1,-1,1,0,1,-1),x) + H(R(1,0,-1,1,
      -1,1),x) + 2*H(R(1,0,-1,1,1,-1),x) + H(R(1,0,1,-1,1,-1),x);
\end{verbatim}
The conversion of combinations of denominators, logarithms and 
polylogarithms in terms of $\Hpl$-functions is usually a straightforward 
exercise in integration by parts, provided that it can be done at all. We 
give one example of an integral
\begin{verbatim}
#-
#include harmpol.h
#call htables(6)
CF  int;
Off Statistics;
.global
Local F = ln_(1+x)^2*ln_(1-x)^2*ln_(x)/x*int(x,0,1);
id  ln_(1+x) =  H(R(-1),x);
id  ln_(1-x) = -H(R(1),x);
id  ln_(x)   =  H(R(0),x);
#call hbasis(H,x)
id  H(R(?a),x)/x*int(x,0,1) = H(R(0,?a),1)-H(R(0,?a),0);
id  H(R(?a,n?{1,-1},?b),0) = 0;  * Provided there is a nonzero index
#do i = 1,6
id  H(R(n1?,...,n`i'?),1) = Htab`i'(n1,...,n`i');
#enddo
.sort
On Statistics;
Print +f;
.end
\end{verbatim}
which gives
\begin{verbatim}
    #-

Time =       0.61 sec    Generated terms =         10
                F        Terms in output =         10
                         Bytes used      =        172

   F = - 1/2*z2*ln2^4 - 129/140*z2^3 + 7/6*z3*ln2^3 - 37/16*z3^2
       - 31/8*z5*ln2 + 8*ln2*li5half + 4*ln2^2*li4half
       + 1/9*ln2^6 + 8*li6half + 2*s6;
\end{verbatim}
The constants are explained both in the paper and the htable files. These 
integrals are not easy by conventional techniques.

\subsection{Meltran}

This package does the Mellin transforms and inverse Mellin transforms 
between the harmonic polylogarithms and the harmonic sums. It allows 
calculations to be done in either space and then converted to the other, 
whatever is easiest. The definition of a Mellin transform is:
\begin{eqnarray}
 M(f(x),N) & = & \int_0^1dx\ x^N f(x) \nonumber \\
 M(\frac{f(x)}{(1\minus x)_+},N) & = & \int_0^1dx\ \frac{x^N 
       f(x)-f(1)}{1\minus x} \nonumber \\
 M(\frac{f(x)\ln^p(1\minus x)}{(1\minus x)_+},N) & = &
   \int_0^1dx\ \frac{\left(x^N 
       f(x)-f(1)\right)\ln^p(1\minus x)}{1\minus x}
\end{eqnarray}
in which the function $f$ is supposed to be finite for $x=1$ when the 
factor $1/(1\minus x)_+$ is present. Hence we see that the first step in 
the Mellin transformation of a $\Hpl$-function is the extraction of 
potential powers of $\ln(1-x)$. The Mellin transform of a $\Hpl$-function, 
divided by either $1-x$ or $1+x$ can be solved recursively and the result 
is a combination of harmonic sums in the Mellin parameter $N$. There is a 
one to one correspondence between the Mellin transform of $\Hpl$-functions 
of weight $w$, divided by either $1-x$ or $1+x$ and harmonic sums of weight 
$w+1$ (the Mellin transform can also contain sums of a lower weight, but of 
the highest weight there is only a single term). Hence the inverse Mellin 
transform can also be obtained rather easily. The only complicating factor 
is that there will be terms in $x=1$ or $N\rightarrow\infty$. Currently 
these can only be combined into a `minimal' representation if the weights 
are at most 9.

The mellin package is part of the harmpol package. Hence it is sufficient 
to include the file harmpol.h. Example:
\begin{verbatim}
#-
#include harmpol.h
Off statistics;
.global
L   F = S(R(1,2,1,2),N);
#call invmel(S,N,H,x)
.sort
On Statistics;
Print +f;
.end
\end{verbatim}
results in
\begin{verbatim}
    #-

Time =       2.72 sec    Generated terms =          4
                F        Terms in output =          4
                         Bytes used      =        144

   F =
       - 38/35*z2^3 + 1/2*z3^2 - 1/2*H(R(1),x)*[1-x]^-1*z2^2
       - H(R(1,0,1,1,0),x)*[1-x]^-1;
\end{verbatim}

Of course it should be clear that with these techniques one can do many 
integrals involving combinations of powers of $x$, denominators with $1\pm 
x$, $\ln(x)$, $\ln(1\pm x)$ and polylogarithms with a large variety of 
arguments. This is usually a soft spot in most computer algebra systems.

\subsection{Color}

This package handles group invariants. The theory behind this is given in 
ref.~\cite{Color}. All internal variables in this package have a name that 
starts with the characters cOl in order to avoid potential name conflicts. 
With this package one can, for instance, compute the group invariants that 
are needed in complicated Feynman diagram calculations in such a way that 
the values for a specific representation of a specific group can be 
substituted at a later stage. This makes calculations more general. A 
simple example of the use of this package is the calculation of a diagram 
with 14 vertices each having three legs, connected in such a way that the 
smallest loop contains 6 vertices. This diagram is unique. If the vertices 
all belong to the adjoint representation the program would be:
\begin{verbatim}
#include color.h
Off Statistics;
.global
G   g14 = cOlf(cOli1,cOli2,cOli3)*cOlf(cOli1,cOli4,cOli5)
         *cOlf(cOli2,cOli6,cOli7)*cOlf(cOli3,cOli8,cOli9)
         *cOlf(cOli4,cOli10,cOli11)*cOlf(cOli5,cOli12,cOli13)
         *cOlf(cOli6,cOli14,cOli15)*cOlf(cOli7,cOli16,cOli17)
         *cOlf(cOli8,cOli18,cOli19)*cOlf(cOli9,cOli20,cOli21)
         *cOlf(cOli10,cOli21,cOli15)*cOlf(cOli13,cOli19,cOli14)
         *cOlf(cOli17,cOli11,cOli18)*cOlf(cOli12,cOli16,cOli20);
sum cOli1,...,cOli21;
.sort
#call docolor
.sort
On Statistics;
Print +f +s;
.end
\end{verbatim}
and its output is
\begin{verbatim}
Time =       1.93 sec    Generated terms =          3
              g14        Terms in output =          3
                         Bytes used      =         82

   g14 =
       + 1/648*cOlNA*cOlcA^7
       - 8/15*cOld444(cOlpA1,cOlpA2,cOlpA3)*cOlcA
       + 16/9*cOld644(cOlpA1,cOlpA2,cOlpA3)
      ;
\end{verbatim}
This was run on a pentium 300. Note that the customary f has to be given as 
cOlf etc. Similarly the regular $C_A$ is given as cOlcA. The values for the 
invariants for a given group can be found in the paper. As an example we 
give here the value of E8:
\begin{verbatim}
*
*   Numbers for exceptional algebras
*
S   [NA+2],eta;
id  cOld644(cOlpA1,cOlpA2,cOlpA3) = 175/48*cOlcA^7*cOlNA/[NA+2]^2;
id  cOld444(cOlpA1,cOlpA2,cOlpA3) = cOlcA^6*cOlNA/[NA+2]^2
                                    *(125/27+125/216*cOlNA);
*
*   And for E8
*
id  cOlNA = 248;
id  1/[NA+2] = 1/250;
id  cOlcA = 30*eta;
Print +f +s;
.end
\end{verbatim}
which gives
\begin{verbatim}
   g14 =
       + 2075760000*eta^7
      ;
\end{verbatim}
with the value of $\eta$ usually being two.

\subsection{Mincer}

This is a pure particle physics package. It concerns the evaluation of 
massless propagator type integrals with up to three loops. In addition 
there are routines for the expansion in (fixed) Mellin moments. Computation 
time for higher Mellin moments can be rather large. In addition the disk 
requirements increase exponentially with the number of the moment. One 
needs to specify the topology of the diagram and a few other options. A 
simple example is given by
\begin{verbatim}
#-
#define SCHEME "0"
#define TOPO "la"
V   p1,...,p8,Q;
#include mincer.h
Off Statistics;
.global
L   Dia = 4*p7.p8^2/p1.p1/p2.p2/p3.p3/p4.p4^2/p5.p5/p6.p6/p7.p7/p8.p8*Q.Q;
.sort
#call integral(`TOPO')
.sort
On Statistics;
Multiply ep^3;
id  ep = 0;
print;
.end
\end{verbatim}
which results in
\begin{verbatim}
    #-
~~~Answer in MS-bar

Time =       0.85 sec    Generated terms =          5
              Dia        Terms in output =          5
                         Bytes used      =         64

   Dia =
       - 139/2 - 5/2*ep^-3 - 19/2*ep^-2 - 55/2*ep^-1 + 5/2*z3;
\end{verbatim}
For more details one should have a look at the paper that comes with the 
package. This mincer package has been in use by the author for 9 years now. 
Over the years it has steadily been improved and optimized. The code is 
basically code for version 2 of FORM. It is this package that was used to 
test the parallel version.

\end{appendix}


\begin{thebibliography}{10}
\bibitem{schoonschip} Schoonschip is a program, made by M. Veltman. See H. 
Strubbe, Comp. Phys. Comm. {\bf 8}:1-30,1974.
 
\bibitem{Parallel1}
D. Fliegner, A. R\'etey and J.A.M. Vermaseren, TTP-99-15, hep-ph/9906426,
D. Fliegner, A. R\'etey and J.A.M. Vermaseren, TTP-00-12, hep-ph/0007221.

\bibitem{HarmonicSums}
J.A.M. Vermaseren Int. J. Mod. Phys. {\bf A14},2037 (1999)

\bibitem{broadhurst}
L. Euler, Novi Comm. Acad. Sci. Petropol. 20 (1775) 140.
D. Zagier, First European Congress of Mathematics, Volume 
II, Birkh\"auser, Boston, 1994, pp. 497-512.
J.M. Borwein, D.M. Bradley, D.J. Broadhurst, 
hep-th/9611004, The Electronic Journal of Combinatorics, Vol. 4, No. 2
(Wilf Fetschrift), 1997, \#R5 
(http://www.combinatorics.org/Volume\_4/wilftoc.html).

\bibitem{broadhurstprivate} D. Broadhurst, private communication

\bibitem{HarmonicPoly}
E. Remiddi and J.A.M. Vermaseren, Int. J. Mod. Phys. {\bf A15},725 (2000).

\bibitem{Color}
T. van Ritbergen, A. N. Schellekens and J.A.M. Vermaseren
 Int. J. Mod. Phys. {\bf A14},41 (1999).



\end{thebibliography}
\end{document}